\documentclass[lettersize,journal]{IEEEtran}
\usepackage{cite}
\usepackage{amsmath,amssymb,amsfonts}
\usepackage{graphicx}
\usepackage{subcaption}
\usepackage{algorithm}
\usepackage{algpseudocode} 
\usepackage{hyperref}
\usepackage{xcolor}
\usepackage{xspace}
\usepackage{multirow}
\usepackage{caption}
\usepackage{bm}
\usepackage{booktabs}

\newcommand{\M} {\ensuremath{\mathbf{M}}\xspace}
\newcommand{\A} {\ensuremath{\mathbf{A}}\xspace}
\newcommand{\x} {\ensuremath{\mathbf{x}}\xspace}
\newcommand{\y} {\ensuremath{\mathbf{y}}\xspace}
\newcommand{\T} {\ensuremath{\bm{\theta}}\xspace}

\newcommand{\changes}{\textcolor{black}}

\begin{document}
\title{Scan-Adaptive MRI Undersampling
\\
Using Neighbor-based Optimization (SUNO)}
\author{Siddhant Gautam, \IEEEmembership{Student Member, IEEE},
Angqi Li, Nicole Seiberlich,
Jeffrey A. Fessler, \IEEEmembership{Fellow, IEEE},
and Saiprasad Ravishankar, \IEEEmembership{Senior Member, IEEE}

\thanks{S. Gautam and A. Li are with the
Department of Computational Mathematics, Science and Engineering,
Michigan State University, East Lansing, MI 48824 USA
(e-mail: gautamsi@msu.edu; liangqi1@msu.edu).}
\thanks{N. Seiberlich is with the Department of Radiology, University of Michigan,
Ann Arbor, MI 48109 USA (e-mail: nse@med.umich.edu).}
\thanks{J. A. Fessler is with the Department of Electrical Engineering and Computer Science and the Department of Biomedical Engineering, University of Michigan, Ann Arbor, MI 48109 USA (e-mail: fessler@umich.edu).}
\thanks{S. Ravishankar is with the Department of Computational Mathematics, Science and Engineering and the Department of Biomedical Engineering, Michigan State University, East Lansing, MI 48824 USA (e-mail: ravisha3@msu.edu).}
\thanks{This work was partially supported by the National Institutes of Health (NIH) under Grant R21 EB030762.}}

\maketitle


\begin{abstract}
\changes{Accelerated MRI aims to reduce scan time by acquiring data more efficiently, for example, through optimized pulse sequences or readouts that increase $k$-space coverage per excitation (e.g., echo planar imaging), or by collecting partial $k$-space measurements with advanced reconstruction methods. Acceleration via partial $k$-space acquisition (i.e., undersampling) has received significant attention, particularly with the rise of learning-based reconstruction methods.}
\changes{Recent works have explored population-adaptive sampling patterns learned from groups of patients (or scans), which enhance sampling pattern design by tailoring it to dataset-specific characteristics, rather than relying on generic approaches. Building on this idea, sampling techniques can be further personalized down to the level of individual scans, enabling the capture of subject- or slice-specific details that may be overlooked in population-based designs. To address this challenging problem, we propose a framework for jointly learning scan-adaptive Cartesian undersampling patterns and a corresponding reconstruction model from a training set, enabling more tailored sampling for individual scans.}
We use an alternating algorithm for learning the sampling patterns and the reconstruction model
where we use an iterative coordinate descent (ICD)
based offline optimization of scan-adaptive $k$-space sampling patterns
for each example in the training set.
A nearest neighbor search is then used
to select the scan-adaptive sampling pattern at test time
from initially acquired low-frequency $k$-space information.
We applied the proposed framework (dubbed SUNO)
to the fastMRI multi-coil knee and brain datasets,
demonstrating improved performance over the currently used undersampling patterns
at both $4\times$ and $8\times$ acceleration factors
in terms of both visual quality and quantitative metrics.
The code for the proposed framework is available at
\url{https://github.com/sidgautam95/adaptive-sampling-mri-suno}.
\end{abstract}


\section{Introduction}\label{sec:introduction}
Magnetic Resonance Imaging (MRI) is a widely used non-invasive biomedical imaging technology
that allows visualization of both anatomical structures and physiological functions.
Some of its benefits include a lack of ionizing radiation and excellent soft-tissue contrast.
MRI scanners sequentially collect measurements in the time (or spatial frequency) domain (known as $k$-space),
from which an image is reconstructed.
The scanner must sample numerous $k$-space points
in order to estimate an image with a clinically appropriate spatial resolution,
which causes the acquisition process
to be slow and expensive.
Accelerating MRI scans reduces acquisition time,
reduces patient discomfort, increases scaling throughput, and reduces motion artifacts.
Such acceleration often requires choosing an appropriate undersampling pattern or trajectory
along with a reconstruction model that enables accurate recovery from reduced measurements.

{Some of the earliest approaches for accelerating MR imaging included pulse sequence and $k$-space trajectory design~\cite{liang2000principles, bernstein2004handbook, tsao2010ultrafast} and parallel imaging~\cite{pruessmann1999sense, griswold2002generalized, ying2010parallel, deshmane2012parallel}. Parallel imaging exploits coil sensitivity information but is limited by noise amplification and residual artifacts at higher accelerations. Unlike traditional MRI, which follows the Nyquist–Shannon sampling requirement, compressed sensing (CS) allows sub-Nyquist $k$-space sampling by exploiting image sparsity in transform domains for accurate reconstruction~\cite{donoho2006compressed, lustig2007sparse, lustig2008compressed}. Building on this sparsity-driven framework, recent approaches have explored learned image models for reconstruction, including synthesis dictionary learning~\cite{ravishankar2010mr, lingala2013blind, qu2014magnetic, zhan2015fast} and transform learning~\cite{ravishankar2012learning, ravishankar2019image}.}

{Over the years, machine learning approaches have been explored for reconstructing MR images from undersampled measurements, including model-driven methods such as ADMM-Net~\cite{sun2016deep,yang2017admm} and ISTA-Net~\cite{zhang2018ista}. With the advent of deep learning, convolutional neural networks (CNNs) have achieved tremendous success in this task. U-Net architectures~\cite{ronneberger2015u} trained in a supervised manner have been widely applied for artifact removal~\cite{hyun2018deep, zhu2018image, wang2018image, knoll2020deep}. Similarly, variational networks combine neural networks with the MR forward model to address accelerated multi-coil MRI reconstruction~\cite{hammernik2018learning, sriram2020end}. GAN-based methods exploit adversarial learning to improve perceptual quality~\cite{mardani2017deep, yang2017dagan}. More recently, MoDL~\cite{aggarwal2018modl, aggarwal2020j} has become particularly popular, where the MRI forward model is incorporated within a data consistency term and a CNN reconstructor is employed as a learned denoiser to regularize the reconstruction.}

{Alongside these reconstruction methods, the choice of undersampling pattern is also a crucial aspect to consider. Commonly used undersampling patterns in CS-MRI include variable density~\cite{lustig2007sparse}, uniform random~\cite{gamper2008compressed},  equispaced~\cite{haldar2010compressed}, Poisson-disc~\cite{bridson2007fast}, and combined variable density and Poisson-disc~\cite{lustig2010spirit, levine20173d}. Beyond these fixed sampling patterns, early optimization-based approaches aimed to learn undersampling patterns directly from training $k$-space data by minimizing reconstruction error~\cite{ravishankar2011adaptive}.}
Subsequent work on sampling optimization designed adaptive sampling patterns
using the power spectra of the reference $k$-space data~\cite{knoll2011adapted, vellagoundar2015robust} or the energy preserving sampling method~\cite{zhang2014energy}. Statistical experiment design techniques for MRI sampling prediction were proposed that used the Cramer-Rao lower bound~\cite{haldar2019oedipus, seeger2010optimization}. 
Later, the greedy algorithm and its variations were used to learn a single population-adaptive sampling pattern over a training set of images with a specific choice of reconstruction method~\cite{gozcu2018learning, gozcu2019rethinking}.
Since these approaches learn the undersampling pattern using greedy algorithms over a large number of images, the computational cost involved is high, and it scales quadratically with the number of lines in the mask.
To avoid this, a stochastic version of the greedy mask learning algorithm was proposed that resolved the scaling issues of the previous greedy approaches~\cite{sanchez2020scalable}.
{Recently, bias-accelerated subset selection (BASS)~\cite{zibetti2021fast, zibetti2022alternating} was introduced for parallel MRI as a scalable subset-selection method for population-based sampling pattern learning, providing a more efficient alternative to purely greedy approaches.}

{Deep learning approaches have also been proposed} that jointly learn a sampling pattern
and a corresponding trained reconstruction network
~\cite{bahadir2020deep, zhang2020extending, aggarwal2020j, sherry2020learning, zibetti2021fast, yin2021end, zibetti2022alternating, huang2022single, alkan2024autosamp}. 
LOUPE~\cite{bahadir2020deep} and its multi-coil extension~\cite{zhang2020extending} determine the probability of sampling each pixel or row/column in the $k$-space domain,
with the underlying parameters learned jointly with those of the reconstructor (U-Net).
Similarly, J-MoDL~\cite{aggarwal2020j} jointly learns an MoDL reconstruction network
and a sampling pattern whose parameters are optimized separately along the row and column directions.
More recently, AutoSamp~\cite{alkan2024autosamp} was proposed for joint optimization of sampling patterns and reconstruction in 3D MRI using variational information maximization.
These works can be divided into those predicting Cartesian undersampling patterns
~\cite{bahadir2020deep, yin2021end, zibetti2021fast, zibetti2022alternating, huang2022single, alkan2024autosamp}
and those learning non-Cartesian patterns~\cite{aggarwal2020j}.
Other recent works for learning non-Cartesian sampling trajectories include PILOT~\cite{weiss2019pilot},
SPARKLING~\cite{lazarus2019sparkling, chaithya2022optimizing}, BJORK~\cite{wang2022b}, NC-PDNet~\cite{ramzi2022nc}, and SNOPY~\cite{wang2023stochastic}.

Sequential decision processes have also been applied to undersampling prediction,
where sampling patterns are learned sequentially using reinforcement learning.
In these problems, the sampling optimization is formulated
as a partially observable Markov decision process (POMDP)~\cite{pineda2020active, bakker2020experimental}.

One limitation of the more common population-adaptive approaches~\cite{gozcu2018learning, bahadir2020deep, aggarwal2020j, zibetti2021fast, zibetti2022alternating} 
is that they learn a single sampling pattern suited to the entire dataset rather than adapting to individual scans. 
{To address this, scan-adaptive undersampling techniques have been proposed
that generate sampling patterns individually for each scan by leveraging subject-specific anatomy.
For example, SeqMRI~\cite{yin2021end} trains a reconstruction model jointly with a sampler that predicts sampling patterns sequentially, whereas MNet~\cite{huang2022single} jointly trains a reconstruction network and a CNN-based sampler to predict undersampling patterns directly from low-frequency $k$-space data. Such scan-specific approaches can improve reconstruction quality by exploiting key slices or image-specific features, offering potential advantages over conventional population-adaptive techniques.}
The population-adaptive patterns can be learned offline,
whereas scan-specific sampling patterns
must be determined rapidly
while the subject is in the scanner, after collecting some preliminary k-space data.

This paper proposes a framework for jointly learning scan-adaptive 1D Cartesian undersampling patterns
and a reconstruction model for multi-coil MRI on a training dataset.
Our algorithm alternatingly estimates a reconstructor and a collection of sampling patterns from training data. 
We use a sampling optimization algorithm based on iterative coordinate descent
to yield improved sampling patterns on training data
and use the nearest neighbor search to determine such patterns at test time based on acquired low-frequency $k$-space.
{The key methodological difference between the proposed SUNO framework and recent scan-adaptive methods like SeqMRI \cite{yin2021end} (or e.g., M-Net~\cite{huang2022single}) is that the SeqMRI employs a fully differentiable approach, jointly optimizing a sequential sampling policy and a reconstruction strategy via standard backpropagation. In contrast, SUNO formulates mask learning as an integer programming problem and uses a dictionary of learned sampling patterns to find the best mask at test time.} 
Our results show that the scan-adaptive Cartesian sampling patterns yield better reconstruction quality in terms of NRMSE, SSIM~\cite{wang2004image}, and PSNR metrics, compared to existing baselines for multi-coil MRI.
This paper builds upon our previous short conference work~\cite{gautam2024patient} 
and extends it to higher acceleration factors and learns undersampling patterns over different anatomies. We also present extended comparisons with several baselines.

The rest of this paper is organized as follows.
Section~\ref{sec:methods} discusses the details of the MRI forward model, deep learning-based reconstruction,
and the details of our proposed training framework that alternates
between optimizing a reconstructor and updating scan-adaptive sampling patterns on a training set.
Section~\ref{sec:experiments} discusses the details of training datasets and implementation details.
Section~\ref{sec:results} presents the results of applying our approach to the fastMRI dataset
and compares it with existing baselines.
We also provide ablation studies on our sampling pattern optimization algorithm.
Section~\ref{sec:discussion} provides a summary of our findings and possible new directions for future work, and further conclusions are provided in Section~\ref{sec:conclusion}.

\section{Methods}\label{sec:methods}
\subsection{Multi-coil MRI Reconstruction}
In multi-coil MRI reconstruction, the goal is to recover the underlying MR image $\x\in \mathbb{C}^n$ from a set of undersampled multi-coil measurements $ \y \in \mathbb{C}^m$.
The regularized MRI reconstruction problem can be formulated as follows:
\begin{equation}
    \underset{\x}{\min} \, \| \M \A \x-\y\|_2^2 + \lambda \mathcal{R}(\x)
\end{equation}

Here, $\M$ is an operator that subsamples $k$-space,
$\A= \mathbf{F} \mathbf{S}$ is the fully sampled MRI measurement operator
and $\mathcal{R}(\x)$ is a regularizer.
$\mathbf{F}$ is the 2D Fourier transform operator
and $\mathbf{S}$ encodes the sensitivity maps of the receiver coils.
The regularizer $\mathcal{R}(\x)$ typically captures assumed properties of the image
and can take on various forms such as total variation, or low-rank or transform-domain sparsity penalties.

{Recently, deep learning has become an increasingly powerful tool for MRI reconstruction, eliminating the need for hand-crafted regularizers. In this paper, we discuss four such approaches: U-Net~\cite{ronneberger2015u, hyun2018deep}, MoDL~\cite{aggarwal2018modl}, VarNet~\cite{hammernik2018learning, sriram2020end}, and ZS-SSL~\cite{yaman2022zero}.  
U-Net~\cite{ronneberger2015u} is a convolutional encoder–decoder that predicts the underlying clean image directly from the aliased input. Building on this, model-based deep learning (MoDL)~\cite{aggarwal2018modl} adopts a fully unrolled, end-to-end trainable framework that alternates between data consistency enforced via the MR forward model and CNN-based denoising, formulated as  
\begin{equation} \label{eqmodl}
    \hat{\x} = \underset{\x}{\arg\min} \, \| \M \A \x - \y \|_2^2 + \lambda \|\x - D_{\T}(\x)\|_2^2,
\end{equation}
where $D_{\T}(\x)$ is a CNN denoiser with parameters $\T$.  
Variational networks~\cite{hammernik2018learning} follow a similar approach, alternating between the $k$-space data consistency and CNN-based image-domain priors. End-to-end VarNet (E2E-VarNet)~\cite{sriram2020end} further extends this approach by fully unrolling the reconstruction process and incorporating $k$-space domain processing, achieving state-of-the-art performance in multi-coil MRI. Finally, Zero-Shot Self-Supervised Learning (ZS-SSL)~\cite{yaman2022zero} trains a network directly on undersampled measurements from individual scans by partitioning $k$-space into disjoint sets for training, self-supervision, and validation, enabling scan-specific reconstructions without fully sampled references.}

\begin{figure}[ht]
    \centering
    \includegraphics[width=0.9\linewidth]{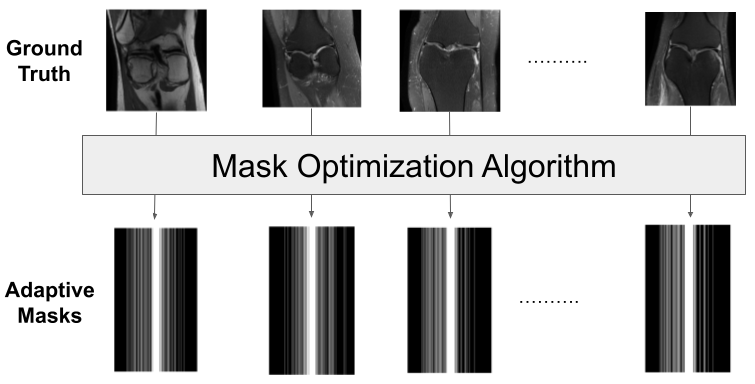}
    \caption{{Illustration of scan-adaptive undersampling patterns generated by our proposed sampling optimization framework. Central $k$-space is consistently sampled across cases, while the high frequency lines vary across the subjects, reflecting individual anatomy and coil sensitivity profiles.}}
    \label{fig:mri_train_pipeline}
\end{figure}

{In this paper,
we focus on learning scan-specific sampling patterns $\{ \M_i \}$ instead of a single population-adaptive sampling pattern.}
In this framework, we first optimize these scan-adaptive masks $\{ \M_i \}$ offline for the training set. Then at test time, these masks are chosen using a nearest neighbor search, as described in a later subsection. {Figure~\ref{fig:mri_train_pipeline} shows the scan adaptive sampling patterns learned by the proposed sampling optimization framework.
The optimized patterns consistently sample
the $k$-space center across subjects,
whereas the high-frequency lines vary depending on the anatomy of the scan
and the coil sensitivities, reflecting the scan-adaptive nature of the approach.}

\subsection{Framework for Jointly Learning Reconstructor and Sampler} \label{sec:alternate_framework}

This section presents our proposed approach
for jointly learning a set of scan-adaptive Cartesian undersampling patterns
$\{ \M_i \}$
along with a reconstructor trained
to be suitable for all of these undersampling patterns. 
Using a training set consisting of fully sampled $k$-space and corresponding ground truth images, we learn a collection of scan-adaptive sampling masks and a reconstructor from the training data.
We formulated the joint optimization problem as follows:
\begin{equation}
    \underset{\T, \, \M_i \in \mathcal{C},\,i\in\{1,\cdots,N\}}{\min}
    \sum_{i=1}^N \| f_{\T} (\A_i^H \M_i \y^{\text{full}}_i ) - \x^{\text{gt}}_i \|_2^2,
    \label{eq:alternate_min}
\end{equation}
where $\M_i \in \mathcal{C}$ is the $i$th training $k$-space subsampling mask that inserts zeros at non-sampled locations, $\y^{\text{full}}_i$ and $\x^{\text{gt}}_i$ are the $i$th fully sampled multi-coil training $k$-space and the corresponding ground truth image, respectively 
and $N$ is the number of training images. The set $\mathcal{C}$ denotes all the 1D Cartesian undersampling patterns with a specified sampling budget. $\A_i^H$ is the adjoint of the fully sampled multi-coil MRI measurement operator for the $i$th training scan, and $f_{\T}$ is the reconstruction network trained on the set of sampling patterns
$\{ \M_i \}$. 

\begin{figure}[ht]
    \centering
    \includegraphics[width=0.9\linewidth]{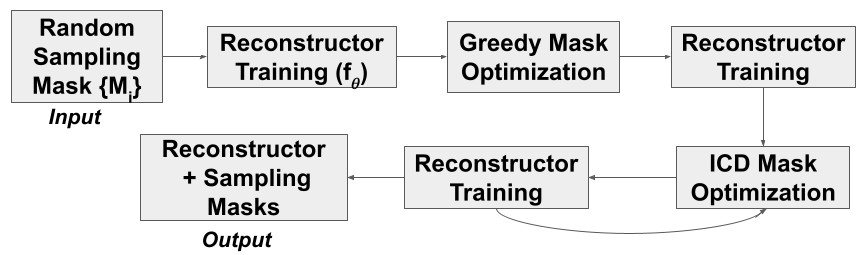}
    \caption{Alternating framework for mask and reconstructor update during joint training. The first four steps serve to create a good initialization for the mask and reconstructor optimization. The masks could be alternatively initialized with, e.g., population-adapted patterns.}
     \label{fig:maintraining}
\end{figure} 

We use the alternating framework shown in Figure~\ref{fig:maintraining}
to solve this highly challenging optimization problem.
The algorithm starts with {random masks} as an initial guess~\cite{lustig2007sparse}
and alternates between updating a reconstructor and sampling masks until we get a final set of scan-adaptive masks $\{ \M_i \}$
and a reconstruction network $f_{\T}$ trained on them.  
For optimizing the scan-adaptive masks,
we initially use a greedy~\cite{gozcu2018learning}
and later our proposed ICD-based sampling optimization algorithm.
More details of the sampling optimization algorithm are in the next section.

\subsection{Iterative Coordinate Descent (ICD) based Sampling Optimization} \label{sec:gicd_algorithm}

\begin{figure}[ht]
    \centering
    \includegraphics[width=0.9\linewidth]{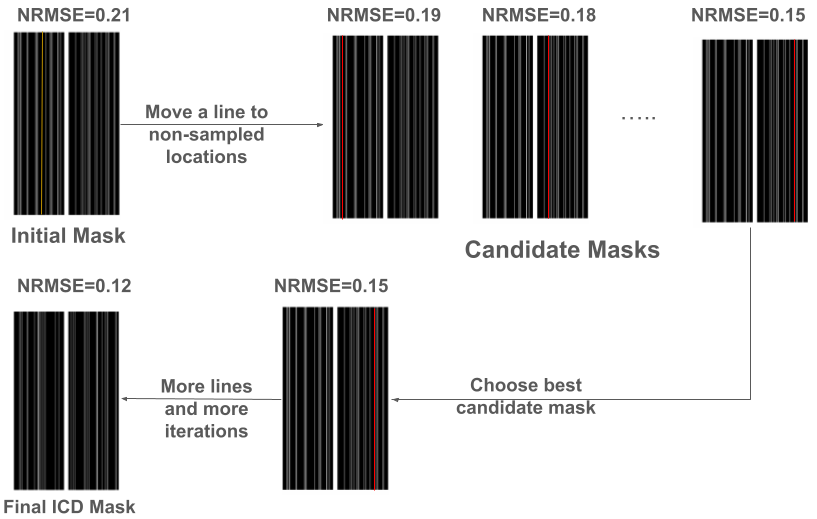}
    \caption{Schematic of offline iterative coordinate descent (ICD) based sampling pattern optimization.}
    \label{fig:icd_flow}
\end{figure}

A greedy algorithm was proposed in prior work~\cite{gozcu2018learning} 
to optimize high-quality sampling patterns that specify samples in $k$-space
that minimize the reconstruction error given a choice of the reconstruction model used.
Starting with no sampled lines or only fixed low-frequency lines,
at each step of the greedy sampling optimization,
the $k$-space phase encoding line that gives the lowest reconstruction error is added to a particular sampling mask.
The algorithm keeps adding lines until the sampling budget is reached. However, the sampling pattern obtained using the greedy algorithm can be sub-optimal and can be further optimized using an iterative coordinate descent (ICD) based sampling optimization. 
The proposed iterative coordinate descent (ICD) based sampling optimization algorithm further optimizes the greedy mask iteratively by picking one line at a time in the current mask and moving it to the best new location in terms of the reconstruction error, and cycling over all lines to move in this manner. Figure~\ref{fig:icd_flow} shows the schematic of mask updates during various steps of the ICD sampling optimization. The steps of the algorithm are given in detail in Algorithm~\ref{alg:reg_icd}. Thus, the ICD sampling optimization further improves the greedy masks and yields better quality scan-adaptive masks. The optimized masks depend on the choice of the initial mask, the reconstructor used, and the metric chosen for the loss function.
The ICD sampling optimization algorithm ensures a monotonic decrease and convergence of the non-negative reconstruction loss~\eqref{eq:alternate_min}.

\begin{algorithm}[!t]
\caption{Sampling Pattern Optimization}
\label{alg::mb}
\begin{algorithmic}[1]
\Require Fully sampled $k$-space $\y^{\text{full}}$ and corresponding forward operator $\A$, ground truth image $\x^{\text{gt}}$, reconstructor $f$,  loss function $L$, budget $B$, number of ICD iterations $N_{\text{iter}}$,  set of all possible line locations $\mathbf{S}$, set of locations of initial sampled lines $\mathbf{\Omega}_{\mathrm{initial}}$, initial mask $\M_{\mathbf{\Omega}_{\mathrm{initial}}}$
\State $\mathbf{\Omega} \leftarrow \mathbf{\Omega}_{initial}$
\For{$j=1:N_{\text{iter}}$}
\State $\{l_i\}_{i=1}^B\leftarrow$ entries in current $\mathbf{\Omega}$

\For{$i=1:B$}
    \State $\mathbf{\Omega}' = \mathbf{\Omega}  \setminus  l_i$ 
    \State $\mathbf{\Omega} \leftarrow \mathbf{\Omega}' \cup S^*$ where
    \begin{equation*}
        S^*=\underset{S \in \mathbf{S},\, S \notin \mathbf{\Omega}'}{\arg\min} 
        \, L(\x^{\text{gt}},f(\A^H \M_{\mathbf{\Omega}' \cup S} \y^{\text{full}}))
    \end{equation*}
    where $\M_{\mathbf{\Omega}' \cup S}$ is the operator sampling along lines at $\mathbf{\Omega}' \cup S$.
\EndFor
\EndFor
\State \textbf{return} $\mathbf{\Omega}$
\end{algorithmic} \label{alg:reg_icd}
\end{algorithm}

\subsection{Neighbor based Sampling Prediction} \label{sec:nn_search}
This subsection describes our approach to predict the sampling pattern
from initially acquired $k$-space measurements at testing time.
Given our set of scan-adaptive sampling patterns obtained from the training process, the task at test time is to estimate the high-frequency lines in $k$-space based on initially acquired low-frequency information.
We use the nearest neighbor search to predict the sampling pattern from the collection of training scans.
The nearest neighbor is found by comparing the adjoint reconstruction of the low-frequency test $k$-space
and the corresponding low-frequency part of the training $k$-space as follows:

\begin{equation}
   d_i =  d(\A^H_{\mathrm{test}} \y^{\mathrm{lf}}_{\mathrm{test}},
   \A^H_{\mathrm{train_i}} \y^{\mathrm{lf}}_{\mathrm{train_i}}),
   \label{eq:nn}
\end{equation}
where $\y^{\mathrm{lf}}_{\mathrm{test}}$ and $\y^{\mathrm{lf}}_{\mathrm{train_i}}$
are the low-frequency part of testing and training $k$-space with zeros at high frequencies.
$\A^H_{\mathrm{test}}$ and $\A^H_{\mathrm{train_i}}$
are the adjoints of the fully sampled MRI forward operators
for the test and $i$th training scans, respectively.
Different metrics $d$ can be used to define the nearest neighbors,
e.g., Euclidean distance, structural similarity index (SSIM)~\cite{wang2004image}, or normalized cross-correlation. 
\changes{In this work, we used the Euclidean distance
between the zero-filled reconstructions obtained from the low-frequency $k$-space data
to identify the nearest neighbor.}
We choose the optimized mask of the nearest neighbor (called the SUNO mask)
and use that at test time in the scanner to collect the rest of the measurements.

\begin{figure*}[ht]
    \centering
    \includegraphics[width=0.9\linewidth]{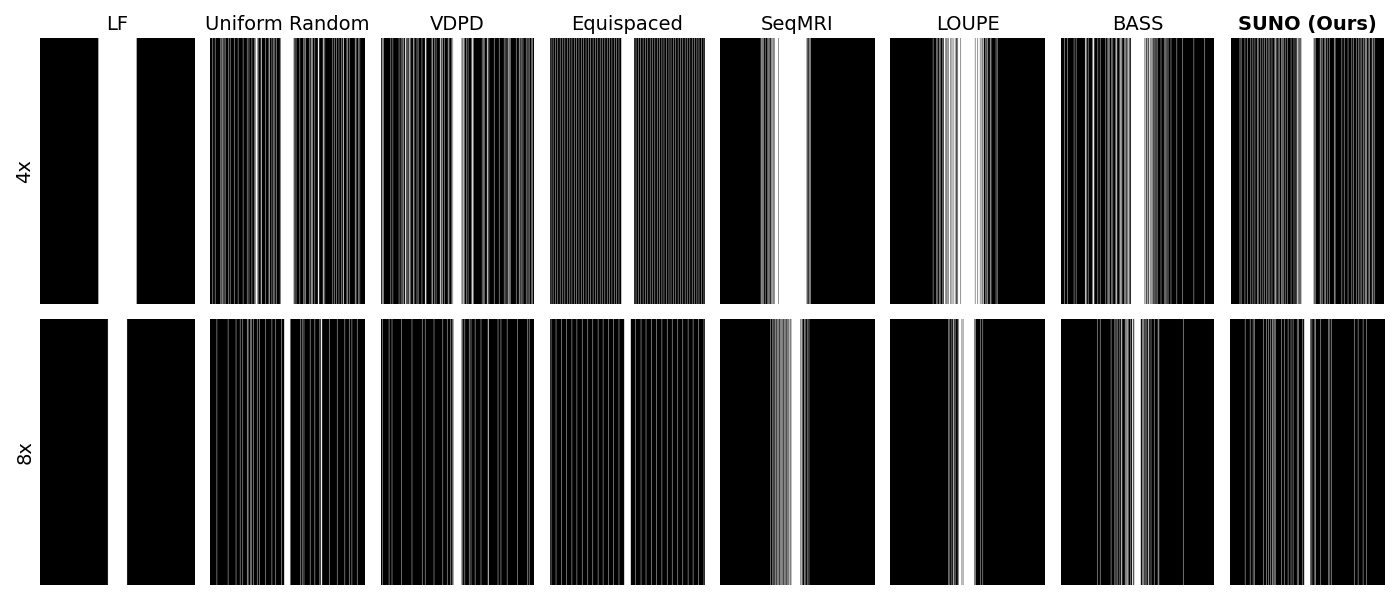}
    \caption{Comparison of different masks used for reconstruction at a) $4\times$ and b) $8\times$ acceleration factors for the knee dataset. Masks displayed are: 1) low-frequency (fixed), 2) 1D Uniform Random (random), 3) VDPD (random), 4) Equispaced (fixed), 5) SeqMRI (optimized - scan adaptive), 6) LOUPE (optimized - population adaptive), 7) BASS (optimized - population adaptive), and 8) SUNO (optimized - scan adaptive). {For the scan-adaptive methods (SeqMRI and SUNO), only one representative mask is shown here for each acceleration factor; additional instances illustrating variability are provided in the supplementary document.}}
\label{fig:baseline_masks}
\end{figure*}

\section{Experiments}\label{sec:experiments}
\subsection{Datasets}
Our experiments used the fastMRI multi-coil knee and brain datasets~\cite{zbontar2018fastmri, knoll2020fastmri}. 
The details for each dataset are as follows:

\subsubsection{fastMRI Multi-coil Knee Dataset}
The fastMRI multi-coil knee dataset contains images collected using two different pulse sequences, yielding coronal proton-density weighted images with (PDFS) and without (PD) fat suppression. For our experiments, we used a total of 156 scans (comprising both PD and PDFS scans) and split them into training, validation, and testing sets.
From each scan, we discarded the first 10 and last 5 slices due to a lack of identifiable image features, which gave us 1514, 194, and 104 training, validation, and testing slices, respectively.
Each image was collected using 15 coils of $k$-space data with a matrix size of $640\times 368$. We used the ESPIRIT calibration approach~\cite{pruessmann1999sense,uecker2014espirit} to estimate the sensitivity maps from the central 30 lines of $k$-space. 

\subsubsection{fastMRI Multi-coil Brain Dataset}
To test the generalization of our proposed sampling prediction algorithm, we also applied our algorithm on the fastMRI multi-coil brain dataset which consists of FLAIR, T1-weighted, and T2-weighted images. 
From this dataset, we used a total of 1660 slices for our experiments and split them into 1480, 120, and 60 training, validation, and testing images, respectively. The scans were acquired with a matrix acquisition size of $640 \times 320$, and the number of receiver coils varied between 4 and 20 across different scans. The sensitivity maps were estimated using the ESPIRiT calibration approach.

\subsection{Comparison with Other Undersampling Patterns}\label{sec:baseline_masks}

{We compared our proposed SUNO sampling patterns with several baselines. 
Classical baselines included low-frequency (LF), uniform random~\cite{gamper2008compressed}, equispaced~\cite{haldar2010compressed}, and variable-density Poisson-disc (VDPD)~\cite{bridson2007fast, levine20173d} masks. 
Learned baselines included scan-adaptive SeqMRI~\cite{yin2021end}, population-adaptive LOUPE~\cite{bahadir2020deep}, and BASS~\cite{zibetti2021fast, zibetti2022alternating} (all Cartesian). All methods were adapted for 1D Cartesian sampling along the phase-encoding ($k_y$) direction for clinical 2D MRI.}

For SeqMRI and LOUPE, we used their publicly available PyTorch implementations%
\footnote{\url{https://github.com/tianweiy/SeqMRI};
\\\url{https://github.com/cagladbahadir/LOUPE}}
and extended them to the multi-coil setting. In LOUPE, the slope parameter was set to $\alpha=5$, with a learning rate of $10^{-3}$ for mask updates. SeqMRI was trained with a learning rate of \(5 \times 10^{-5}\), halved every 10 epochs, with four sequential steps and an SSIM loss between real-valued magnitude-only images. For reconstruction, LOUPE used a two-channel U-Net, while SeqMRI used a two-channel residual U-Net; both networks started with 64 channels and were trained for 100 epochs. For each undersampling pattern, 30 and 15 central low-frequency lines were fixed for $4\times$ and $8\times$ acceleration, respectively.
{In our experiments, the learned LOUPE masks at $4\times$ and $8\times$ accelerations showed different sampling distributions compared to their prior work~\cite{bahadir2020deep}. These differences can be partly due to our use of a multi-coil MRI setup with a larger matrix size ($640\times368$) compared to their original single-coil study with a smaller matrix size ($320\times320$).
Additionally, in our implementation, we incorporated a straight-through (ST) estimator~\cite{bengio2013estimating} for binary mask generation during training, similar to the approach used in the multi-coil extension of LOUPE~\cite{zhang2020extending}.}

{For obtaining the BASS mask, we used the MATLAB implementation available at the authors’ website.\footnote{\url{https://cai2r.net/resources/data-driven-learning-of-mri-sampling-pattern/}} 
We followed the default settings in the released code, and used the sampling pattern (SP) learning parameter $\alpha = 0.9$ in the 1D SP mode. As with the other baselines, $K_{\text{LF}}$ central lines were fixed for calibration for each acceleration factor. The algorithm was run for $L = 100$ iterations for both cases. Similar to the LOUPE setting, the BASS mask was optimized over the whole training set. Separate masks were optimized on the brain dataset at acceleration factors of 4$\times$ and 8$\times$.}

\changes{All reconstruction models (ZS-SSL, E2E-VarNet, MoDL) were retrained separately for every mask setting, including population-adaptive masks such as BASS, to ensure fair and consistent comparisons across all methods.}
Figure~\ref{fig:baseline_masks} shows representative SUNO and baseline masks for the knee dataset.
{Here we adapted the VDPD approach~\cite{levine20173d} to 1D Cartesian undersampling along the phase-encoding direction
for a fair comparison with our learned masks. Although the VDPD masks are adapted for 1D Cartesian undersampling along the phase-encoding direction, they still follow the Poisson-disc principle with enforced minimum spacing. In 1D, the variable-density pattern can visually resemble a uniform random mask, but the underlying sampling distribution is different; these differences become more apparent in 2D sampling.}

Table~\ref{tab:icd_param} lists the parameters used in the sampling optimization algorithm for each acceleration factor. The number of ICD iterations $N_{iter}$ was set to 1, as most of the loss reduction occurs in the first iteration, while additional iterations provide marginal improvements at higher computational cost.

\begin{table}[ht]
\centering
\begin{tabular}{lcc}
\toprule
\textbf{Acceleration Factor}   & $\mathbf{4\times}$ & $\mathbf{8\times}$ \\
\midrule
Total sampled lines ($B$)                & 92  & 46  \\
Centrally fixed lines ($c$)              & 30  & 15  \\
Lines updated by ICD ($m=B-c$)               & 62  & 31  \\
Candidate search space ($N_y-B$)       & 276 & 322 \\
No. of ICD iterations ($N_{iter}$)              & 1   & 1   \\
\bottomrule
\end{tabular}
\caption{Parameters involved in the ICD sampling optimization algorithm with $k$-space dimension $N_x \times N_y$, where $N_x = 640$ and $N_y = 368$, for the fastMRI multi-coil knee dataset. One-third of the lines are fixed at the center of $k$-space ($c$), and the remaining lines ($m$) are optimized by the algorithm. The candidate search space for each update is $(N_y - B)$. For the brain dataset, the parameters scale accordingly with the dataset matrix size.}
\label{tab:icd_param}
\end{table}

\subsection{Implementation Details}
Our algorithms were implemented in Python, using the PyTorch package.
We used two-channel reconstruction networks to obtain the underlying image from the undersampled $k$-space,
with the two channels being the real and imaginary parts of the complex image.
We used Facebook Research's implementation of U-Net in the PyTorch framework\footnote{\url{https://github.com/facebookresearch/fastMRI/blob/main/fastmri/models/unet.py}}.
For MoDL~\cite{aggarwal2018modl}, we used a deep iterative up-down (DIDN) network~\cite{yu2019deep} as the denoiser inside the training framework.
We used 6 unrollings of the denoiser and the conjugate gradient (CG) block.
The regularization parameter $\lambda$ controlling the weighting of the two terms
(see Eq.\eqref{eqmodl}) was set to $10^2$
and the tolerance for the CG algorithm used was $10^{-5}$ after tuning them on multiple images.
{
For VarNet~\cite{sriram2020end}, we used 12 cascades, each containing a U-Net with 18 channels in the first layer, using the official implementation provided by Facebook Research.\footnote{\url{https://github.com/facebookresearch/fastMRI/blob/main/fastmri/models/varnet.py}} Sensitivity maps used for VarNet training were estimated using the ESPIRIT calibration approach~\cite{uecker2014espirit}. The reconstructed image was obtained by applying the adjoint of the forward MRI operator to the reconstructed $k$-space from VarNet. 
Both MoDL and VarNet were trained for 100 epochs with a batch size of 1. 
Adam optimizer~\cite{kingma2014adam} was used for training the networks with a learning rate of $10^{-3}$.  
For the ZS-SSL approach~\cite{yaman2022zero}, we employed 10 unrolled blocks and 15 residual blocks, along with 10 conjugate gradient iterations for data consistency. The model was trained for 300 epochs with a learning rate of $5 \times 10^{-4}$ using the official implementation available on GitHub.\footnote{\url{https://github.com/byaman14/ZS-SSL-PyTorch}}
}
The simulations were performed on an NVIDIA RTX A5000 GPU with 24 GB RAM.


\begin{figure*}[!ht]
    \centering
\includegraphics[width=0.9\linewidth]{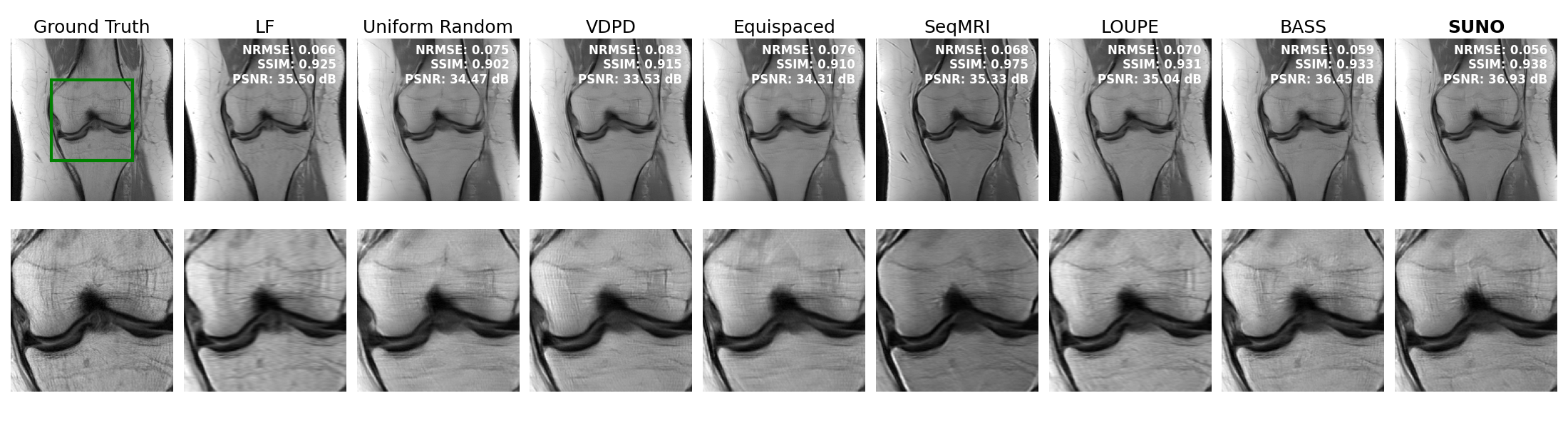}
        \caption{Reconstructed images using MoDL reconstruction network at $4\times$ acceleration factor for a testing slice. The second row shows the zoom-in images from the area inside the green rectangle.
        \changes{The SUNO approach outperformed the rest in terms of visual quality and better preserved structural detail, whereas BASS offered similar performance.}}
        \label{fig:icd_recon_knee_4x}
\end{figure*}

\begin{figure*}[!ht]
    \centering
\includegraphics[width=0.9\linewidth]{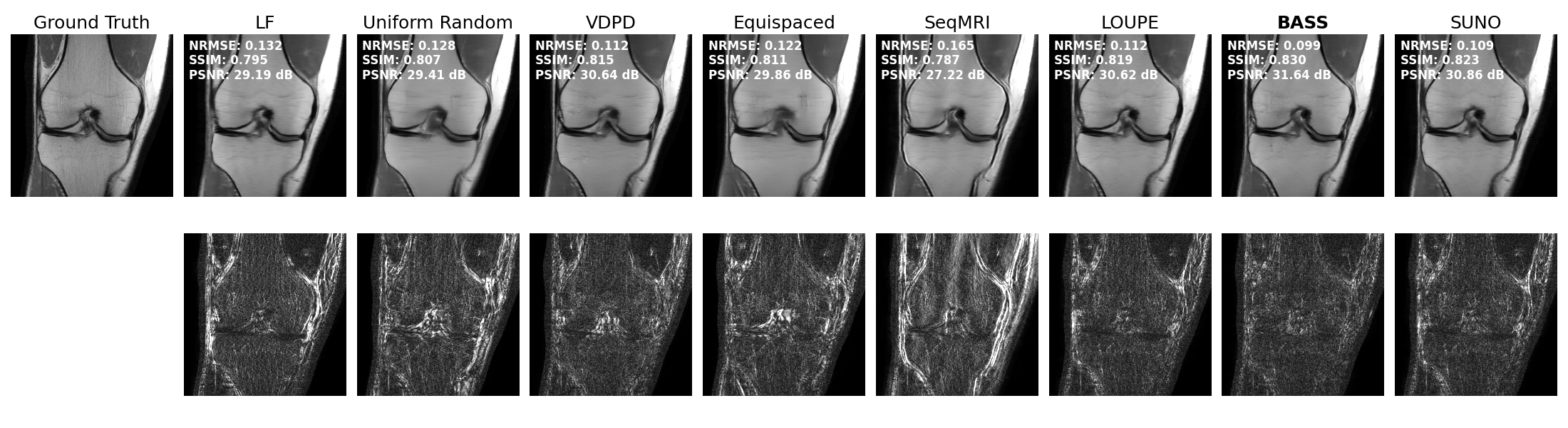}
    \caption{{Reconstructed images using MoDL reconstruction network at $8\times$ acceleration factor with the error maps in the second row. At 8$\times$, the population-adaptive BASS mask achieved the lowest reconstruction error and best preserved structural details, \changes{while SUNO still provided competitive reconstructions.} The error images (magnitudes) are shown in the range $[0,\,0.1]$.}}
    \label{fig:icd_recon_knee_8x}
\end{figure*}

\subsection{Performance Metrics}
To evaluate the quality of reconstructed images, we used normalized root mean squared error (NRMSE),
structural similarity index (SSIM)~\cite{wang2004image},
and peak signal-to-noise ratio (PSNR) as the metrics.
These metrics assess the similarity between the ground truth and the reconstructed images.
Lower NRMSE, higher SSIM, and PSNR values suggest better reconstruction quality.
All the metrics were evaluated on the central $320\times 320$ portion of the image.




\section{Results}\label{sec:results}
\subsection{Studies on the fastMRI Multi-Coil Knee Dataset}

In this section, we show the result of applying our optimized scan-adaptive SUNO masks on the fastMRI multi-coil knee dataset at $4\times$ and $8\times$ acceleration factors. We compare the quality of the reconstructed images using our optimized masks with the other baseline masks described in Section~\ref{sec:baseline_masks}. 

Figure~\ref{fig:icd_recon_knee_4x} shows the {images reconstructed using the MoDL network} (along with zoom-ins over a region of interest) at $4\times$ acceleration. It is clear from the figure that the proposed \changes{SUNO provides} better-reconstructed images compared to other baselines in terms of NRMSE, SSIM, and PSNR metrics. The zoom-ins also show that the reconstructed image using the SUNO mask preserves fine structural details present in the ground truth. Figure~\ref{fig:icd_recon_knee_8x} shows reconstructed and error images at $8\times$ acceleration using the \changes{SUNO} along with other baselines.
The error maps indicate that BASS yields the best quality reconstruction, with \changes{SUNO} giving comparable reconstructions.

Table~\ref{tab:metrics_knee} reports the mean and standard deviation values of NRMSE, SSIM, and PSNR for reconstructed images using ZS-SSL~\cite{yaman2022zero}, E2E-VarNet~\cite{sriram2020end}, and MoDL~\cite{aggarwal2018modl}.
On average, the proposed \changes{SUNO} outperformed most baselines across different acceleration factors and reconstructors.
At $4\times$ acceleration, \changes{SUNO achieved the best performance with both VarNet and MoDL}.
However, at $8\times$ acceleration, the BASS mask performed comparably to, and in some cases slightly better than \changes{SUNO} for VarNet and MoDL.
For the ZS-SSL reconstruction method, \changes{SUNO} gave the best performance in terms of the reconstruction metrics
at both 4$\times$ and 8$\times$ acceleration factors.



\begin{table*}[ht]
\centering
\begin{tabular}{l l ccc ccc}
\toprule
& & \multicolumn{3}{c}{4$\times$} & \multicolumn{3}{c}{8$\times$} \\
\cmidrule(lr){3-5} \cmidrule(lr){6-8}
Reconstructor & Mask & NRMSE $\downarrow$ & SSIM $\uparrow$ & PSNR (dB) $\uparrow$ & NRMSE $\downarrow$ & SSIM $\uparrow$ & PSNR (dB) $\uparrow$ \\
\midrule
\multirow{8}{*}{\centering {ZS-SSL}~\cite{yaman2022zero}} & 
LF & \multicolumn{1}{c|}{0.156 $\pm$ 0.074} & \multicolumn{1}{c|}{0.794 $\pm$ 0.072} & 29.15 $\pm$ 2.80 
& \multicolumn{1}{c|}{0.260 $\pm$ 0.100} & \multicolumn{1}{c|}{0.690 $\pm$ 0.090} & 24.90 $\pm$ 3.20 \\ \cline{2-8}
& Uniform Random & \multicolumn{1}{c|}{0.125 $\pm$ 0.052} & \multicolumn{1}{c|}{0.834 $\pm$ 0.074} & 30.94 $\pm$ 2.87  
& \multicolumn{1}{c|}{0.235 $\pm$ 0.095} & \multicolumn{1}{c|}{0.700 $\pm$ 0.085} & 25.50 $\pm$ 3.00 \\ \cline{2-8}
& {VDPD}~\cite{bridson2007fast, levine20173d} & \multicolumn{1}{c|}{0.123 $\pm$ 0.060} & \multicolumn{1}{c|}{0.832 $\pm$ 0.083} & 31.30 $\pm$ 3.40  
& \multicolumn{1}{c|}{0.200 $\pm$ 0.095} & \multicolumn{1}{c|}{0.730 $\pm$ 0.095} & 27.20 $\pm$ 3.00 \\ \cline{2-8}
& Equispaced~\cite{haldar2010compressed} & \multicolumn{1}{c|}{0.117 $\pm$ 0.053} & \multicolumn{1}{c|}{0.841 $\pm$ 0.070} & 31.62 $\pm$ 2.94 
& \multicolumn{1}{c|}{0.232 $\pm$ 0.107} & \multicolumn{1}{c|}{0.705 $\pm$ 0.085} & 25.64 $\pm$ 2.82 \\ \cline{2-8}
& SeqMRI~\cite{yin2021end} & \multicolumn{1}{c|}{0.118 $\pm$ 0.040} & \multicolumn{1}{c|}{0.846 $\pm$ 0.081} & 31.56 $\pm$ 2.89 
& \multicolumn{1}{c|}{0.199 $\pm$ 0.113} & \multicolumn{1}{c|}{0.732 $\pm$ 0.118} & 27.12 $\pm$ 3.42 \\ \cline{2-8}
& LOUPE~\cite{bahadir2020deep} & \multicolumn{1}{c|}{0.119 $\pm$ 0.057} & \multicolumn{1}{c|}{0.848 $\pm$ 0.071} & 31.54 $\pm$ 3.22  
& \multicolumn{1}{c|}{0.193 $\pm$ 0.112} & \multicolumn{1}{c|}{0.746 $\pm$ 0.102} & 27.67 $\pm$ 3.48 \\ \cline{2-8}
& {BASS}~\cite{zibetti2021fast, zibetti2022alternating} & \multicolumn{1}{c|}{0.113 $\pm$ 0.060} & \multicolumn{1}{c|}{0.840 $\pm$ 0.089} & 32.20 $\pm$ 3.40  
& \multicolumn{1}{c|}{0.195 $\pm$ 0.100} & \multicolumn{1}{c|}{0.740 $\pm$ 0.090} & 27.30 $\pm$ 3.10 \\ \cline{2-8}
& \textbf{SUNO (Ours)} & \multicolumn{1}{c|}{\textbf{0.110 $\pm$ 0.058}} & \multicolumn{1}{c|}{\textbf{0.860 $\pm$ 0.081}} & \textbf{32.38 $\pm$ 3.43}  
& \multicolumn{1}{c|}{\textbf{0.192 $\pm$ 0.116}} & \multicolumn{1}{c|}{\textbf{0.746 $\pm$ 0.104}} & \textbf{27.69 $\pm$ 3.50} \\ \hline

\multirow{8}{*}{\centering {E2E-VarNet}~\cite{sriram2020end}} 
& LF & \multicolumn{1}{c|}{0.133 $\pm$ 0.059} & \multicolumn{1}{c|}{0.866 $\pm$ 0.052} & 30.50 $\pm$ 2.77  
& \multicolumn{1}{c|}{0.194 $\pm$ 0.114} & \multicolumn{1}{c|}{0.783 $\pm$ 0.071} & 27.55 $\pm$ 3.07 \\ \cline{2-8}

& Uniform Random & \multicolumn{1}{c|}{0.127 $\pm$ 0.053} & \multicolumn{1}{c|}{0.864 $\pm$ 0.055} & 30.79 $\pm$ 2.74  
& \multicolumn{1}{c|}{0.186 $\pm$ 0.064} & \multicolumn{1}{c|}{0.801 $\pm$ 0.065} & 27.67 $\pm$ 2.49 \\ \cline{2-8}

& {VDPD}~\cite{bridson2007fast, levine20173d} & \multicolumn{1}{c|}{0.121 $\pm$ 0.050} & \multicolumn{1}{c|}{0.869 $\pm$ 0.055} & 31.25 $\pm$ 2.78  
& \multicolumn{1}{c|}{0.151 $\pm$ 0.060} & \multicolumn{1}{c|}{0.817 $\pm$ 0.068} & 29.31 $\pm$ 2.63 \\ \cline{2-8}

& Equispaced~\cite{haldar2010compressed} & \multicolumn{1}{c|}{0.118 $\pm$ 0.046} & \multicolumn{1}{c|}{0.873 $\pm$ 0.054} & 31.56 $\pm$ 2.76  
& \multicolumn{1}{c|}{0.182 $\pm$ 0.061} & \multicolumn{1}{c|}{0.812 $\pm$ 0.059} & 27.86 $\pm$ 2.33 \\ \cline{2-8}

& SeqMRI~\cite{yin2021end} & \multicolumn{1}{c|}{0.117 $\pm$ 0.052} & \multicolumn{1}{c|}{0.892 $\pm$ 0.052} & 31.58 $\pm$ 2.81  
& \multicolumn{1}{c|}{0.155 $\pm$ 0.054} & \multicolumn{1}{c|}{0.833 $\pm$ 0.062} & 28.90 $\pm$ 2.35 \\ \cline{2-8}

& LOUPE~\cite{bahadir2020deep} & \multicolumn{1}{c|}{0.113 $\pm$ 0.051} & \multicolumn{1}{c|}{0.890 $\pm$ 0.051} & 32.00 $\pm$ 2.90  
& \multicolumn{1}{c|}{0.148 $\pm$ 0.064} & \multicolumn{1}{c|}{0.828 $\pm$ 0.065} & 29.51 $\pm$ 2.66 \\ \cline{2-8}

& {BASS}~\cite{zibetti2021fast, zibetti2022alternating} & \multicolumn{1}{c|}{0.108 $\pm$ 0.050} & \multicolumn{1}{c|}{0.890 $\pm$ 0.053} & 32.48 $\pm$ 2.99  
& \multicolumn{1}{c|}{\textbf{0.140 $\pm$ 0.061}} & \multicolumn{1}{c|}{\textbf{0.843 $\pm$ 0.063}} & \textbf{29.99 $\pm$ 2.62} \\ \cline{2-8}

& \changes{\textbf{SUNO (Ours)}} & \multicolumn{1}{c|}{\textbf{0.107 $\pm$ 0.051}} & \multicolumn{1}{c|}{\textbf{0.896 $\pm$ 0.053}} & \textbf{{32.55 $\pm$ 3.11}}  
& \multicolumn{1}{c|}{{0.147 $\pm$ 0.062}} & \multicolumn{1}{c|}{{0.828 $\pm$ 0.065}} & {29.57 $\pm$ 2.62} \\
\hline

\multirow{9}{*}{\centering MoDL~\cite{aggarwal2018modl}} 
& LF & \multicolumn{1}{c|}{0.134 $\pm$ 0.066} & \multicolumn{1}{c|}{0.929 $\pm$ 0.031} & 30.59 $\pm$ 3.05 
& \multicolumn{1}{c|}{0.187 $\pm$ 0.094} & \multicolumn{1}{c|}{0.781 $\pm$ 0.071} & 27.67 $\pm$ 2.83 \\ \cline{2-8}
& Uniform Random & \multicolumn{1}{c|}{0.137 $\pm$ 0.051} & \multicolumn{1}{c|}{0.920 $\pm$ 0.031} & 31.08 $\pm$ 2.50 
& \multicolumn{1}{c|}{0.198 $\pm$ 0.065} & \multicolumn{1}{c|}{0.759 $\pm$ 0.068} & 26.76 $\pm$ 2.26 \\ \cline{2-8}
& {VDPD}~\cite{bridson2007fast, levine20173d} & \multicolumn{1}{c|}{0.131 $\pm$ 0.049} & \multicolumn{1}{c|}{0.846 $\pm$ 0.055} & 30.45 $\pm$ 2.85 
& \multicolumn{1}{c|}{0.164 $\pm$ 0.064} & \multicolumn{1}{c|}{0.788 $\pm$ 0.070} & 28.49 $\pm$ 2.63 \\ \cline{2-8}
& Equispaced~\cite{haldar2010compressed} & \multicolumn{1}{c|}{0.127 $\pm$ 0.050} & \multicolumn{1}{c|}{0.927 $\pm$ 0.031} & 30.67 $\pm$ 2.58 
& \multicolumn{1}{c|}{0.190 $\pm$ 0.067} & \multicolumn{1}{c|}{0.766 $\pm$ 0.070} & 27.56 $\pm$ 2.27 \\ \cline{2-8}
& {SeqMRI}~\cite{yin2021end} & \multicolumn{1}{c|}{0.118 $\pm$ 0.048} & \multicolumn{1}{c|}{0.939 $\pm$ 0.028} & 31.50 $\pm$ 2.79 
& \multicolumn{1}{c|}{0.170 $\pm$ 0.058} & \multicolumn{1}{c|}{0.790 $\pm$ 0.062} & 26.30 $\pm$ 2.09 \\ \cline{2-8}
& LOUPE~\cite{bahadir2020deep} & \multicolumn{1}{c|}{0.116 $\pm$ 0.049} & \multicolumn{1}{c|}{0.938 $\pm$ 0.028} & 31.63 $\pm$ 2.81 
& \multicolumn{1}{c|}{0.163 $\pm$ 0.072} & \multicolumn{1}{c|}{0.804 $\pm$ 0.067} & 28.56 $\pm$ 2.78 \\ \cline{2-8}
& {BASS}~\cite{zibetti2021fast, zibetti2022alternating} & \multicolumn{1}{c|}{{0.115 $\pm$ 0.049}} & \multicolumn{1}{c|}{0.867 $\pm$ 0.055} & {31.67 $\pm$ 2.80} 
& \multicolumn{1}{c|}{\textbf{0.155 $\pm$ 0.061}} & \multicolumn{1}{c|}{\textbf{0.805 $\pm$ 0.065}} & \textbf{28.99 $\pm$ 2.44} \\ \cline{2-8}
& \changes{\textbf{SUNO (Ours)}} & \multicolumn{1}{c|}{\textbf{0.114 $\pm$ 0.046}} & \multicolumn{1}{c|}{\textbf{0.940 $\pm$ 0.029}} & \textbf{31.74 $\pm$ 2.85} 
& \multicolumn{1}{c|}{0.162 $\pm$ 0.065} & \multicolumn{1}{c|}{0.801 $\pm$ 0.068} & 28.66 $\pm$ 2.61 \\
\hline
\end{tabular}
\caption{{Distribution of NRMSE, SSIM, and PSNR values for reconstructed images from the multicoil knee dataset at 4$\times$ and 8$\times$ acceleration factors. 
For ZS-SSL, SUNO achieved the best performance at both 4$\times$ and 8$\times$ acceleration. For E2E-VarNet and MoDL, \changes{SUNO} performed best at 4$\times$ acceleration.
At 8$\times$ acceleration, the population-adaptive BASS mask achieved the lowest NRMSE and highest SSIM for E2E-VarNet and MoDL, indicating that a single global mask can provide stability under extreme undersampling, while \changes{SUNO} remained competitive.}}
\label{tab:metrics_knee}
\end{table*}

\subsection{Applicability to Different Anatomies}
To test the applicability of our proposed scan-adaptive sampling prediction approach on different anatomies, we also optimized masks using the proposed training pipeline on the fastMRI multi-coil brain dataset.
Then, using the nearest neighbor search, the masks were predicted at test time, and the performance of these learned SUNO masks was compared with the other baselines - low-frequency, Uniform Random, equispaced, {SeqMRI}, and LOUPE masks.

\begin{figure*}[!ht]
    \centering
    \includegraphics[width=0.9\linewidth]{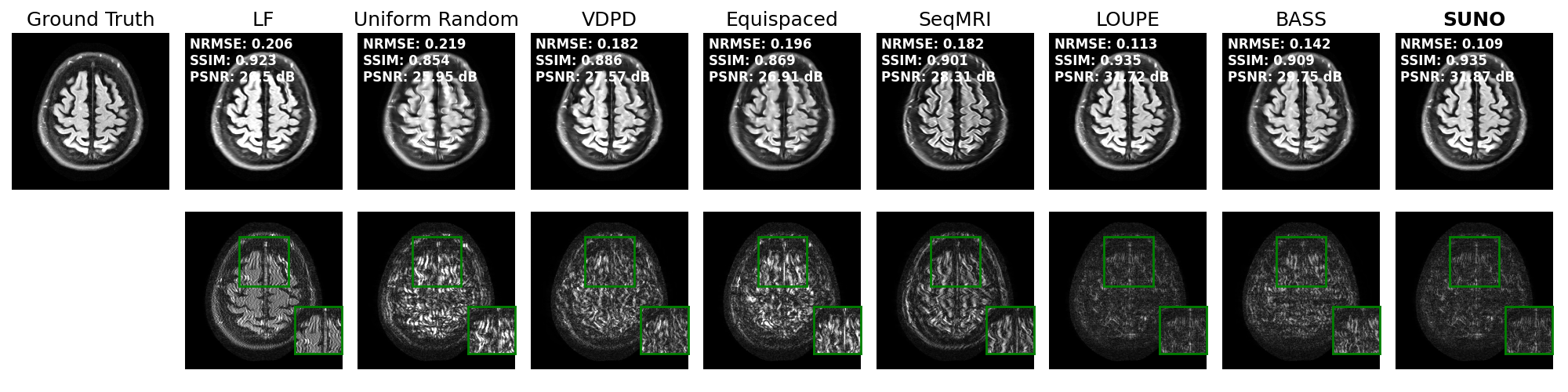}
    \caption{Reconstructed and error images using different undersampling patterns with the MoDL reconstruction network (two-channel) on the fastMRI brain dataset at $4\times$ acceleration. The green rectangle highlights the zoomed-in regions in the error images. The proposed SUNO achieved the best performance in terms of NRMSE and PSNR, while obtaining SSIM comparable to LOUPE.}
    \label{fig:icd_recon_brain_4x}
\end{figure*}

\begin{figure*}[!ht]
    \centering
    \includegraphics[width=0.9\linewidth]{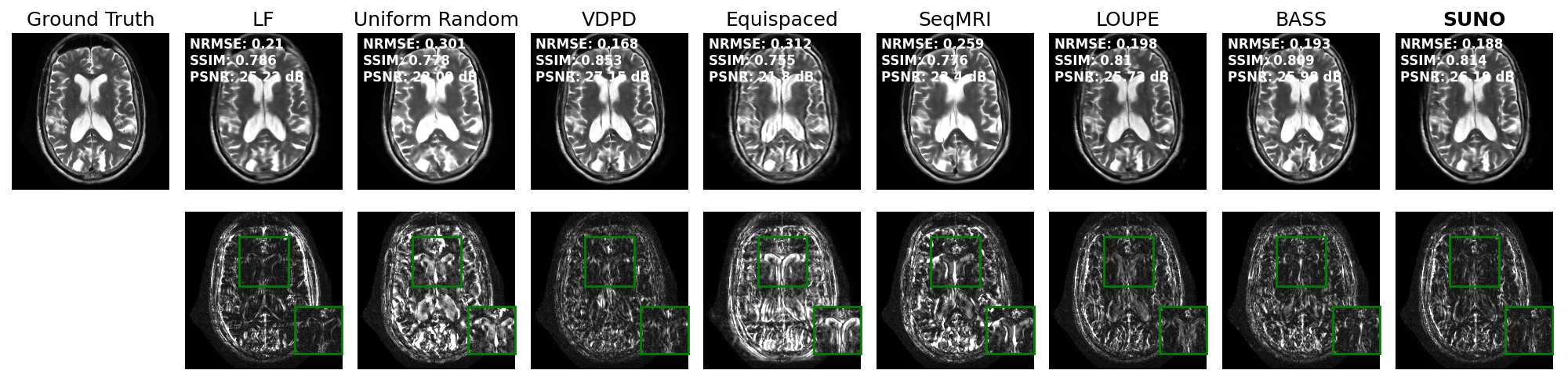}
    \caption{Reconstructed and error images using different undersampling patterns using MoDL reconstruction network(two-channel) on fastMRI brain dataset at $8\times$ acceleration factor. The green rectangle shows the zoomed-in portions in the error image. The proposed SUNO performed better than other baselines in terms of NRMSE, SSIM, and PSNR metrics.}
    \label{fig:icd_recon_brain_8x}
\end{figure*}

Figure~\ref{fig:icd_recon_brain_4x} and~\ref{fig:icd_recon_brain_8x} show the reconstructed and error images using a brain testing slice for $4\times$ and $8\times$ acceleration factors. The figure shows that the optimized scan adaptive SUNO masks outperform the other baseline masks in terms {of the reconstruction metrics} for both $4\times$ and $8\times$ acceleration factors. The error images indicate a lower reconstruction error for SUNO approach.
The mean and standard deviation values of the reconstruction metrics {using ZS-SSL, VarNet, and MoDL reconstruction methods} learned over all test cases are mentioned in Table~\ref{tab:metrics_brain}.
From the table, we can see that the proposed
\changes{SUNO} approach outperformed all baselines for all three reconstruction approaches in terms of the error metrics used.

\begin{table*}[ht]
\centering
\begin{tabular}{l l ccc ccc}
\toprule
& & \multicolumn{3}{c}{4$\times$} & \multicolumn{3}{c}{8$\times$} \\
\cmidrule(lr){3-5} \cmidrule(lr){6-8}
Reconstructor & Mask & NRMSE $\downarrow$ & SSIM $\uparrow$ & PSNR (dB) $\uparrow$ & NRMSE $\downarrow$ & SSIM $\uparrow$ & PSNR (dB) $\uparrow$ \\
\midrule
\multirow{6}{*}{\centering {ZS-SSL}~\cite{yaman2022zero}} & LF & \multicolumn{1}{c|}{0.173 $\pm$ 0.057} & \multicolumn{1}{c|}{0.854 $\pm$ 0.025} & 29.14 $\pm$ 2.46  & \multicolumn{1}{c|}{0.259 $\pm$ 0.067} & \multicolumn{1}{c|}{0.764 $\pm$ 0.046} &  25.45 $\pm$ 2.14 \\ \cline{2-8}
& Uniform Random & \multicolumn{1}{c|}{0.124 $\pm$ 0.034} & \multicolumn{1}{c|}{0.873 $\pm$ 0.029} & 31.83 $\pm$ 2.50  & \multicolumn{1}{c|}{0.241 $\pm$ 0.065} & \multicolumn{1}{c|}{0.802 $\pm$ 0.061} & 26.07 $\pm$ 2.37 \\ \cline{2-8}
& {VDPD}~\cite{bridson2007fast, levine20173d} & \multicolumn{1}{c|}{0.123 $\pm$ 0.033} & \multicolumn{1}{c|}{0.896 $\pm$ 0.027} & 32.02 $\pm$ 2.40  & \multicolumn{1}{c|}{0.247 $\pm$ 0.056} & \multicolumn{1}{c|}{0.840 $\pm$ 0.027} & 25.59 $\pm$ 1.17  \\ \cline{2-8}
& Equispaced~\cite{haldar2010compressed} & \multicolumn{1}{c|}{0.120 $\pm$ 0.032} & \multicolumn{1}{c|}{0.891 $\pm$ 0.028} & 32.13 $\pm$ 2.30 & \multicolumn{1}{c|}{0.297 $\pm$ 0.053} & \multicolumn{1}{c|}{0.742 $\pm$ 0.049} & 24.09 $\pm$ 1.36  \\ \cline{2-8}
& {SeqMRI}~\cite{yin2021end} & \multicolumn{1}{c|}{0.118 $\pm$ 0.027} & \multicolumn{1}{c|}{0.897 $\pm$ 0.025} & 32.22 $\pm$ 2.76  & \multicolumn{1}{c|}{0.171 $\pm$ 0.060} & \multicolumn{1}{c|}{0.841 $\pm$ 0.029} & 27.98 $\pm$ 2.06  \\ \cline{2-8}
& LOUPE~\cite{bahadir2020deep} & \multicolumn{1}{c|}{0.114 $\pm$ 0.024}  & \multicolumn{1}{c|}{0.900 $\pm$ 0.021} & 32.61 $\pm$ 2.20  & \multicolumn{1}{c|}{0.166 $\pm$ 0.037} & \multicolumn{1}{c|}{0.864 $\pm$ 0.025} & 28.94 $\pm$ 1.93  \\ \cline{2-8}
& {BASS}~\cite{zibetti2021fast, zibetti2022alternating} & \multicolumn{1}{c|}{0.119 $\pm$ 0.031} & \multicolumn{1}{c|}{0.892 $\pm$ 0.026} & 32.18 $\pm$ 2.18  
& \multicolumn{1}{c|}{0.161 $\pm$ 0.057} & \multicolumn{1}{c|}{0.864 $\pm$ 0.029} 
& 29.12
$\pm$ 1.93  \\ \cline{2-8}
& \textbf{SUNO (Ours)} & \multicolumn{1}{c|}{\textbf{0.111 $\pm$ 0.025}} & \multicolumn{1}{c|}{\textbf{0.903 $\pm$ 0.020}} & \textbf{32.83 $\pm$ 2.02}  & \multicolumn{1}{c|}{\textbf{0.158 $\pm$ 0.040}} & \multicolumn{1}{c|}{\textbf{0.869 $\pm$ 0.024}} & \textbf{29.35 $\pm$ 1.92 } \\ \hline
\multirow{9}{*}{\centering {E2E-VarNet}~\cite{sriram2020end}} 
& LF & \multicolumn{1}{c|}{0.144 $\pm$ 0.047} & \multicolumn{1}{c|}{0.964 $\pm$ 0.012} & 30.74 $\pm$ 2.63 & \multicolumn{1}{c|}{0.216 $\pm$ 0.064} & \multicolumn{1}{c|}{0.881 $\pm$ 0.035} & 27.03 $\pm$ 2.38 \\ \cline{2-8}

& Uniform Random & \multicolumn{1}{c|}{0.116 $\pm$ 0.025} & \multicolumn{1}{c|}{0.970 $\pm$ 0.010} & 32.37 $\pm$ 1.89 & \multicolumn{1}{c|}{0.196 $\pm$ 0.040} & \multicolumn{1}{c|}{0.889 $\pm$ 0.029} & 27.72 $\pm$ 1.73 \\ \cline{2-8}

& {VDPD}~\cite{bridson2007fast, levine20173d} & \multicolumn{1}{c|}{0.110 $\pm$ 0.021} & \multicolumn{1}{c|}{0.946 $\pm$ 0.013} & 32.58 $\pm$ 1.66 & \multicolumn{1}{c|}{0.208 $\pm$ 0.043} & \multicolumn{1}{c|}{0.880 $\pm$ 0.026} & 27.18 $\pm$ 1.74 \\ \cline{2-8}

& Equispaced~\cite{haldar2010compressed} & \multicolumn{1}{c|}{0.118 $\pm$ 0.022} & \multicolumn{1}{c|}{0.970 $\pm$ 0.009} & 32.20 $\pm$ 1.64 & \multicolumn{1}{c|}{0.171 $\pm$ 0.044} & \multicolumn{1}{c|}{0.923 $\pm$ 0.021} & 28.92 $\pm$ 2.12 \\ \cline{2-8}

& SeqMRI~\cite{yin2021end} & \multicolumn{1}{c|}{0.109 $\pm$ 0.026} & \multicolumn{1}{c|}{0.973 $\pm$ 0.008} & 33.01 $\pm$ 2.09 & \multicolumn{1}{c|}{0.172 $\pm$ 0.044} & \multicolumn{1}{c|}{{0.920 $\pm$ 0.019}} & 28.91 $\pm$ 1.94 \\ \cline{2-8}

& LOUPE~\cite{bahadir2020deep} & \multicolumn{1}{c|}{0.104 $\pm$ 0.024} & \multicolumn{1}{c|}{0.978 $\pm$ 0.006} & 33.40 $\pm$ 2.09 & \multicolumn{1}{c|}{0.165 $\pm$ 0.037} & \multicolumn{1}{c|}{0.910 $\pm$ 0.022} & 29.24 $\pm$ 1.63 \\ \cline{2-8}
& {BASS}~\cite{zibetti2021fast, zibetti2022alternating} 
& \multicolumn{1}{c|}{0.115 $\pm$ 0.023} 
& \multicolumn{1}{c|}{0.958 $\pm$ 0.012} 
& 33.35 $\pm$ 2.12 
& \multicolumn{1}{c|}{0.167 $\pm$ 0.040} 
& \multicolumn{1}{c|}{0.908 $\pm$ 0.017} 
& 29.15 $\pm$ 2.10 \\ \cline{2-8}

& \changes{\textbf{SUNO (Ours)}} & \multicolumn{1}{c|}{\textbf{0.102 $\pm$ 0.023}} & \multicolumn{1}{c|}{\textbf{0.978 $\pm$ 0.006}} & \textbf{33.51 $\pm$ 2.09} & \multicolumn{1}{c|}{\textbf{0.162 $\pm$ 0.044}} & \multicolumn{1}{c|}{\textbf{0.922 $\pm$ 0.020}} & \textbf{29.51 $\pm$ 2.08} \\
\hline
\multirow{6}{*}{\centering MoDL~\cite{aggarwal2018modl}} 
& LF & \multicolumn{1}{c|}{0.154 $\pm$ 0.046} & \multicolumn{1}{c|}{0.956 $\pm$ 0.014} & 30.22 $\pm$ 2.57 
& \multicolumn{1}{c|}{0.239 $\pm$ 0.073} & \multicolumn{1}{c|}{0.843 $\pm$ 0.036} & 26.17 $\pm$ 2.21 \\ \cline{2-8}

& Uniform Random & \multicolumn{1}{c|}{0.175 $\pm$ 0.052} & \multicolumn{1}{c|}{0.938 $\pm$ 0.019} & 29.28 $\pm$ 2.20 
& \multicolumn{1}{c|}{0.252 $\pm$ 0.058} & \multicolumn{1}{c|}{0.827 $\pm$ 0.036} & 25.57 $\pm$ 1.75 \\ \cline{2-8}

& {VDPD}~\cite{bridson2007fast, levine20173d} & \multicolumn{1}{c|}{0.158 $\pm$ 0.036} & \multicolumn{1}{c|}{0.896 $\pm$ 0.018} & 29.77 $\pm$ 2.05 
& \multicolumn{1}{c|}{0.250 $\pm$ 0.055} & \multicolumn{1}{c|}{0.830 $\pm$ 0.034}  & 25.70 $\pm$ 1.78 \\ \cline{2-8}

& Equispaced~\cite{haldar2010compressed} & \multicolumn{1}{c|}{0.156 $\pm$ 0.038} & \multicolumn{1}{c|}{0.941 $\pm$ 0.020} & 29.93 $\pm$ 2.01 
& \multicolumn{1}{c|}{0.257 $\pm$ 0.056} & \multicolumn{1}{c|}{0.823 $\pm$ 0.035} & 25.38 $\pm$ 1.84 \\ \cline{2-8}

& SeqMRI~\cite{yin2021end} & \multicolumn{1}{c|}{0.122 $\pm$ 0.035} & \multicolumn{1}{c|}{0.958 $\pm$ 0.017} & 31.87 $\pm$ 2.15   
& \multicolumn{1}{c|}{0.204 $\pm$ 0.068} & \multicolumn{1}{c|}{0.828 $\pm$ 0.034} & 27.34 $\pm$ 1.57 \\ \cline{2-8}

& LOUPE~\cite{bahadir2020deep} & \multicolumn{1}{c|}{0.119 $\pm$ 0.033} & \multicolumn{1}{c|}{0.931 $\pm$ 0.014} & 32.38 $\pm$ 2.26 
& \multicolumn{1}{c|}{0.192 $\pm$ 0.055} & \multicolumn{1}{c|}{0.870 $\pm$ 0.029} & {28.12 $\pm$ 2.07} \\ \cline{2-8}

& {BASS}~\cite{zibetti2021fast, zibetti2022alternating} & \multicolumn{1}{c|}{0.124 $\pm$ 0.031} & \multicolumn{1}{c|}{0.921 $\pm$ 0.025} & 31.92 $\pm$ 2.20 
& \multicolumn{1}{c|}{0.202 $\pm$ 0.057} & \multicolumn{1}{c|}{0.864 $\pm$ 0.029} & 27.58 $\pm$ 1.93 \\ \cline{2-8}

& \changes{\textbf{SUNO (Ours)}} & \multicolumn{1}{c|}{\textbf{0.117 $\pm$ 0.031}} & \multicolumn{1}{c|}{\textbf{0.962 $\pm$ 0.013}} & \textbf{32.54 $\pm$ 2.20} 
& \multicolumn{1}{c|}{\textbf{0.191 $\pm$ 0.058}} & \multicolumn{1}{c|}{\textbf{0.871 $\pm$ 0.029}} & \textbf{28.16 $\pm$ 2.11} \\ 
\hline
\end{tabular}
\caption{Distribution of NRMSE, SSIM, and PSNR values for the reconstructed images from the multicoil brain dataset at 4$\times$ and 8$\times$ acceleration factors using various masks and reconstructors. 
The SUNO outperforms the rest at both acceleration factors for all three reconstruction methods used.
The values displayed are mean $\pm$ std.}
\label{tab:metrics_brain}
\end{table*}

\subsection{Comparison with the Oracle Case}

In this section, we compare the performance of oracle masks optimized directly on the test slices (using the sampling optimization with a fixed reconstructor) with the ones predicted from the nearest neighbor search (SUNO masks).
Table~\ref{tab:oracle_vs_suno} gives a comparison of the oracle and SUNO masks for the fastMRI knee dataset. We observe that the oracle-optimized masks perform better than the SUNO mask for both acceleration factors, as expected. This is because the oracle mask was optimized for the particular test scan (scan-adaptive) while the SUNO mask uses the mask optimized on the nearest neighbor training scan. Hence, the oracle masks perform slightly better in general. However, we want to emphasize that estimating the oracle masks requires access to the ground truth, making it infeasible at test time.

\begin{table}[ht]
\centering
\begin{tabular}{lcc}
\toprule
Acceleration Factor & $4\times$ & $8\times$ \\
\midrule
Oracle & \textbf{0.111 / 0.942 / 31.87} & \textbf{0.149 / 0.902 / 29.60} \\
SUNO   & 0.114 / 0.940 / 31.74 & 0.162 / 0.801 / 28.66 \\
\bottomrule
\end{tabular}
\caption{\small Comparison of the oracle and SUNO masks on the fastMRI knee dataset test cases. Values are reported as mean NRMSE / SSIM / PSNR.}
\label{tab:oracle_vs_suno}
\end{table}

\subsection{\changes{Effect of Local Neighbor-Adaptive Reconstruction}}
\label{sec:suno_local}

\changes{In this subsection, we evaluate the effect of combining the proposed scan-adaptive SUNO approach with a local neighbor-adaptive reconstruction technique~\cite{liang2024adaptive}.
Adaptive local network training on a small set of nearest neighbors has been shown to improve reconstruction quality compared to globally trained models, particularly when the local training data closely resemble the test image~\cite{liang2024adaptive}. By focusing on a small, relevant subset of scans, the reconstruction network can better adapt to shared image features while avoiding the variability present in large, heterogeneous training sets. }

\changes{We refer to this setting as SUNO-Local, which uses local neighbor-adaptive reconstruction, and compare it with SUNO-Global, where reconstruction is performed using a globally trained MoDL network. For both settings, the SUNO mask for each test scan is selected using the nearest-neighbor search described in Section~\ref{sec:methods}.
}

\changes{For the local training configuration, the MoDL reconstruction network is initialized from the globally trained model and fine-tuned using the $30$ nearest-neighbor training scans for $50$ epochs. A reduced learning rate of $10^{-6}$ is used to mitigate overfitting due to the limited size of the local training set. All the $k$-space data used for local training are undersampled using the SUNO mask predicted for the corresponding test scan, ensuring consistency between the sampling pattern and the locally adapted reconstructor.}

\changes{Table~\ref{tab:suno_local_vs_global} shows the reconstruction metrics for SUNO-Global and SUNO-Local on the fastMRI multi-coil knee dataset at $4\times$ and $8\times$ acceleration using the MoDL reconstructor. We observe that SUNO-Local provides additional improvements over SUNO-Global in some cases, particularly at lower acceleration factors, thus indicating the benefit of including scan-adaptive reconstruction combined with scan-adaptive sampling.}

\begin{table*}[ht]
\centering
\begin{tabular}{lccc ccc}
\toprule
& \multicolumn{3}{c}{4$\times$} & \multicolumn{3}{c}{8$\times$} \\
\cmidrule(lr){2-4} \cmidrule(lr){5-7}
Method & NRMSE $\downarrow$ & SSIM $\uparrow$ & PSNR $\uparrow$
& NRMSE $\downarrow$ & SSIM $\uparrow$ & PSNR $\uparrow$ \\
\midrule
SUNO-Global
& 0.114 $\pm$ 0.046 & 0.940 $\pm$ 0.029 & 31.74 $\pm$ 2.85
& 0.162 $\pm$ 0.065 & 0.801 $\pm$ 0.068 & 28.66 $\pm$ 2.61 \\
SUNO-Local
& \textbf{0.110 $\pm$ 0.045} & \textbf{0.941 $\pm$ 0.029} & \textbf{32.07 $\pm$ 2.83}
& \textbf{0.160 $\pm$ 0.066} & \textbf{0.803 $\pm$ 0.067} & \textbf{28.82 $\pm$ 2.66} \\
\bottomrule
\end{tabular}
\caption{ \changes{Comparison of SUNO with globally trained reconstruction (SUNO-Global) and local neighbor-adaptive reconstruction (SUNO-Local) on the fastMRI multi-coil knee dataset using the MoDL reconstructor. Values are reported as mean $\pm$ standard deviation over the test set.}}
\label{tab:suno_local_vs_global}
\end{table*}

\subsection{Lesion-Focused Evaluation with fastMRI+}
\label{sec:pathology_eval}
{To better evaluate reconstruction quality in regions containing pathologies, we used the publicly available fastMRI+ dataset~\cite{zhao2022fastmri+}
that provides expert-labeled bounding boxes for common abnormalities
aligned with the original fastMRI dataset~\cite{zbontar2018fastmri}.
We applied the proposed SUNO masks and baseline masks to the knee $k$-space data
at 4$\times$ and 8$\times$ acceleration factors.
For evaluations, we used the MoDL reconstruction network trained on the original fastMRI dataset~\cite{zbontar2018fastmri}.
We computed NRMSE, SSIM, and PSNR over the full image to quantify reconstruction accuracy and assess signal fidelity across the entire anatomy.
Table~\ref{tab:fastmri_plus_testing} summarizes
the results for 4$\times$ and 8$\times$ acceleration.
The NRMSE of the reconstructed images obtained using all the masks
was averaged over 115 testing images.
SUNO achieved the lowest NRMSE at 4$\times$ acceleration,
outperforming other sampling patterns
and getting an improved reconstruction of knee pathologies and lesions.
At 8$\times$, BASS achieved the best performance on this set, though SUNO remained competitive with lower NRMSE compared to other baselines.}

\begin{table*}[t]
\centering
\begin{tabular}{lcccccccc}
\toprule
\textbf{Acceleration} & LF & Uniform Random & VDPD & Equispaced & SeqMRI & LOUPE & BASS & SUNO-Global \\
\midrule
4$\times$ & 0.144 & 0.152 & 0.147 & 0.149 & 0.128 & 0.128 & 0.130 & \textbf{0.128}\\
8$\times$ & 0.207 & 0.240 & 0.183 & 0.225 & 0.188 & 0.186 & \textbf{0.177} & 0.182\\
\bottomrule
\end{tabular}
\caption{{Mean NRMSE values over the reconstructed images evaluated over the fastMRI+ test set at 4$\times$ and 8$\times$ undersampling. Lower values indicate better reconstruction quality.}}
\label{tab:fastmri_plus_testing}
\end{table*}

\subsection{Ablation Study - Convergence and Choice of Parameters for the Sampling Optimization Algorithm}
In this section, we show the effect of changing different parameters for running the ICD sampling optimization (Algorithm~\ref{alg:reg_icd}) on the optimized SUNO masks for the fastMRI multi-coil knee dataset.

\subsubsection{Effect of initialization}
\begin{figure}[ht]
\centering
\includegraphics[width=0.6\linewidth]{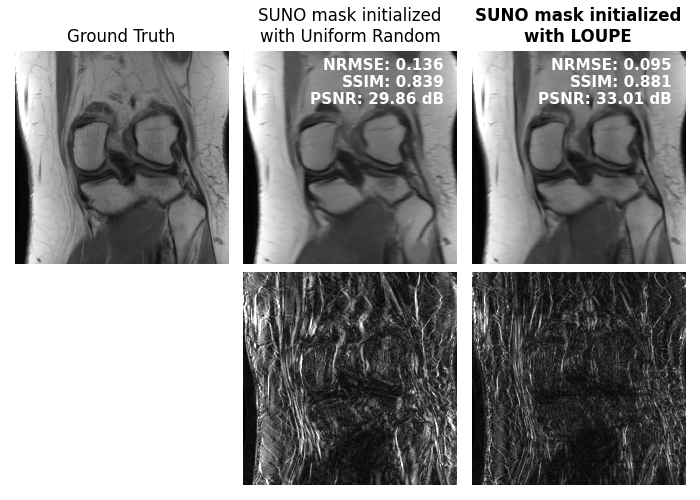}
\caption{\small{Comparing reconstructed images using SUNO masks initialized from {uniform random} and LOUPE masks at $8\times$ acceleration factor.}}
\label{fig:icd_vdrs_vs_loupe}
\end{figure}

In this section, we explore the effect of changing the initial mask used for the sampling pattern optimization - {uniform random} and LOUPE~\cite{bahadir2020deep} masks. The algorithm, when started with a particular mask and given a choice of reconstruction method and loss used, could give different solutions. Table~\ref{tab:icd_vdrs_vs_loupe} lists the performance metrics for the images reconstructed using SUNO masks optimized from a) {uniform random} mask and b) LOUPE mask with the MoDL reconstruction network. Figure~\ref{fig:icd_vdrs_vs_loupe} shows one such example of the reconstructed images obtained from SUNO masks initialized using LOUPE and {uniform random} masks. From the results, we observe that the sampling optimization initialized with the LOUPE mask results in a better reconstruction compared to when it is initialized with a \textbf{uniform random} mask. Since the LOUPE mask is already optimized for multiple training scans (population adaptive), it acts as a better initial point for starting the sampling optimization. The algorithm further optimizes the LOUPE mask for scan-specific details, hence, we get better performance with it compared to LOUPE on test scans. 

\begin{table}[ht]
\centering
\begin{tabular}{lccc}
\toprule
Initial Mask Chosen & NRMSE $\downarrow$ & SSIM $\uparrow$ & PSNR $\uparrow$ \\
\midrule
Uniform Random & 0.164 & 0.896 & 28.45 \\
LOUPE          & \textbf{0.142} & \textbf{0.903} & \textbf{29.78} \\
\bottomrule
\end{tabular}
\caption{Mean reconstruction metrics for masks initialized with uniform random and LOUPE at an $8\times$ acceleration factor, evaluated over 50 test cases. Initializing with LOUPE results in improved reconstruction quality.}
\label{tab:icd_vdrs_vs_loupe}
\end{table}

\subsubsection{Effect of reconstruction method}
\begin{figure}[ht]
    \centering
    \includegraphics[width=0.9\linewidth]{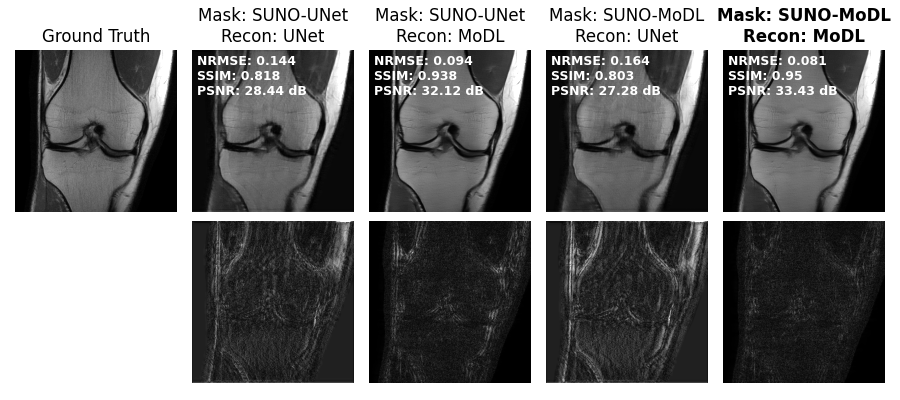}
    \caption{\small{Comparison of reconstructed images from masks optimized using 1) U-Net and 2) MoDL as the reconstruction model.
    For each mask, the reconstructed images using both the U-Net and MoDL networks is shown.  }}
    \label{fig:icd_recon_unet_vs_modl}
\end{figure}

This section shows the effect of the reconstruction method used inside the Algorithm~\ref{alg:reg_icd} on the quality of optimized SUNO masks.
The algorithm works for any choice of reconstruction method,
e.g., compressed sensing (CS) or a pre-trained deep learning model
(e.g., U-Net, MoDL, or E2E-VarNet~\cite{hammernik2018learning, sriram2020end}).
In this paper, we show masks optimized using two such methods - U-Net and MoDL and compare the reconstructed images using these masks.
Figure~\ref{fig:icd_recon_unet_vs_modl} shows the U-Net and MoDL reconstructed images using two optimized masks: one that used U-Net as the reconstruction model in the sampling optimization algorithm and the other with MoDL.
The figure shows that we get the best reconstruction when a better reconstructor (i.e., MoDL network) is used both as the reconstruction model inside the sampling optimization and as the final reconstructor method.

\subsection{Computational Cost of Proposed Approaches} \label{sec:time_complexity}

This section discusses the time complexity of our proposed algorithms:
1) the sampling optimization algorithm and 2) the nearest neighbor search.

\subsubsection{Complexity of Algorithm~\ref{alg:reg_icd}}
{In this section, we evaluate the effect of different parameters on the runtime of our proposed scan-adaptive sampling pattern optimization algorithm. Since our algorithm learns a unique mask for each training scan and slice, an important aspect to consider is the overall computational complexity. This makes training time and resource requirements an important practical consideration, especially for large datasets.}

At each update of ICD, each movable line has $(N_y - B)$ candidate positions, where $N_y$ denotes the number of phase-encoding lines and $B$ is the sampling budget (see Table~\ref{tab:icd_param}). Therefore, across $m$ movable lines (where $m = B - c$, with $c$ being the number of centrally fixed lines) and $N_{iter}$ iterations, the computational cost of ICD sampling optimization is
\[
\mathcal{O}\!\left(N_{\text{iter}}\, m\, (N_y - B)\right),
\]
which is linear in $N_{\text{iter}}$ and approximately linear in $m$ when $N_y \gg m$.
 
Some of the parameters affecting runtime for the Algorithm~\ref{alg:reg_icd} are the reconstruction method that is run repeatedly while moving sampling lines or phase encodes in the mask and the underlying undersampling factor.
Table~\ref{tab:time_icd_vs_recon} shows the dependence of the runtime on both these parameters.
It is clear from the table that the sampling optimization algorithm using a U-Net reconstruction model
results in a lower runtime compared to running the algorithm using a MoDL reconstructor.
This is because the MoDL reconstruction network uses multiple unrollings of the denoiser and the CG block~\cite{aggarwal2018modl}. These empirical observations are consistent with the theoretical scaling described above.

\begin{table}[ht]
\centering
\begin{tabular}{lcc}
\toprule
Reconstructor Used & $4\times$ (min) & $8\times$ (min) \\
\midrule
U-Net & 25.3 & 13.3 \\
MoDL  & 52.1 & 28.6 \\
\bottomrule
\end{tabular}
\caption{Computation time (minutes) for one pass of Algorithm~\ref{alg:reg_icd} (offline) using U-Net and MoDL reconstructors at $4\times$ and $8\times$ acceleration. Experiments were run on an NVIDIA RTX A5000 GPU with 24 GB RAM.}
\label{tab:time_icd_vs_recon}
\end{table}



\subsubsection{Cost of Nearest Neighbor Search}
{In this section, we discuss the time complexity of the nearest neighbor search used for mask selection at test time.
After the initial low-frequency $k$-space lines are acquired, we compute distances between the test scan and all training scans and select the SUNO mask associated with the closest neighbor. The chosen mask is then used to acquire the remaining $k$-space lines, followed by reconstruction with the trained network.}

{Table~\ref{tab:time_vs_neighbors} reports the time required for nearest neighbor mask selection and for a single forward pass of the reconstruction network (inference). These steps introduce only a small computational overhead relative to the overall scan and reconstruction pipeline.}

\begin{table}[ht]
\centering
\begin{tabular}{lc}
\toprule
Procedure & Time (s) \\
\midrule
Nearest Neighbor Mask Selection      & 0.85 \\
Inference                            & 1.53 \\
\bottomrule
\end{tabular}
\caption{{Time taken (in seconds) for nearest neighbor mask selection and reconstruction network inference.}}
\label{tab:time_vs_neighbors}
\end{table}

\section{Discussion}\label{sec:discussion}
We proposed a novel way of learning scan-adaptive Cartesian undersampling patterns for multi-coil MRI. The proposed method demonstrated better accuracy than population-based 1D Cartesian undersampling patterns in terms of NRMSE, SSIM, and PSNR, as well as improved visual quality at 4$\times$ and 8$\times$ acceleration factors. Zoomed-in images highlight improved feature preservation in reconstructions using SUNO masks compared to other baselines. The method was tested on knee and brain datasets, indicating generalization across anatomies.

Similar to the greedy algorithm in prior work~\cite{gozcu2018learning}, the sampling optimization algorithm can be used along with any choice of reconstruction method and the loss metric, giving freedom in designing sampling patterns for different anatomies and different acceleration factors.
{At test time,} a nearest neighbor search was used to predict the pattern from the dictionary of learned patterns.
\changes{Furthermore, we also evaluated the effect of having local neighbor-adaptive reconstruction, which can provide further improvements over a globally trained reconstruction.}


{
The current experiments used a single nearest-neighbor approach for mask selection.
While alternative approaches such as k-nearest neighbor aggregation or learned models could be explored, a systematic evaluation of these techniques was not performed in this work.
The nearest-neighbor approach may be susceptible to outliers as well,
and future research in this direction could explore more robust selection approaches.}

{Our proposed scan-adaptive sampling algorithm is influenced by the choice of initial mask. A population-adaptive mask, such as LOUPE, acts as a better initial point and leads to improved reconstructions compared to a uniform random mask, as the algorithm performs local line-by-line updates rather than large-scale reconfigurations. 
Future work could explore using multiple initializations or randomized perturbations during training to reduce dependence on a single starting mask.}

{We observed that certain small anatomical structures were not well recovered in any of the 8$\times$ accelerated reconstructions, regardless of the method used.
Similarly, in the brain, small structures sometimes appeared distorted or hallucinated in the reconstructed images. These limitations highlight that although the scan-adaptive SUNO masks improve global metrics and preserve most of the anatomical details, recovering some small, low-contrast features remains challenging. 
In the knee, these errors may be due to the need for higher frequencies to capture these small structures, which the learned masks do not fully cover currently. Similarly, in the brain, undersampling combined with the reconstruction network can introduce small spurious features. Optimizing scan-adaptive masks with ROI-specific loss functions may enable better recovery of such fine structures and represents an important direction for future work.}

{In addition to quantitative performance, we analyzed the characteristics of the scan-adaptive undersampling patterns across different subjects. We observed substantial variation in the high-frequency sampling locations across slices and subjects. While part of this variation may arise from random initialization of the masks for each scan, much of it reflects adaptation to individual anatomy and image content.
This combination of shared structure and individualized detail highlights the potential advantages of scan-adaptive sampling over fixed, population-based designs. At the same time, the broader question of whether scan-adaptive sampling is inherently superior to population-based sampling techniques remains open in the deep learning setting. For instance, under certain reconstruction objectives and regularization choices, population-based methods such as LOUPE can outperform some scan-adaptive approaches. To address this, we included recent state-of-the-art human-designed (VDPD) and population-based (BASS) baselines, and our results show that SUNO consistently outperformed them in most cases. Nevertheless, broader benchmarking across architectures and datasets will be needed to fully resolve this important question.}

{While our experiments used 1D undersampling adaptations of VDPD and BASS to remain consistent with the 2D Cartesian acquisitions studied here, we acknowledge that fully 2D undersampling implementations of these methods for 3D acquisitions may provide further insights.
Extending SUNO and such baselines beyond 1D undersampling constitutes an important direction for future work. Though the larger search space in higher dimensions makes these extensions more computationally demanding, this could potentially be alleviated using randomized search techniques.
Recent work, such as AutoSamp~\cite{alkan2024autosamp}, has advanced the design of population-based sampling techniques through more sophisticated optimization and learning formulations. Adapting and comparing such approaches within the Cartesian acquisition setting studied here would help place scan-adaptive methods like SUNO in a broader context of sampling design.}

A drawback of the current method is the time-consuming process of learning scan-adaptive sampling patterns on the whole training set.
More work is required to make the optimization process efficient and faster, to make this approach more feasible. However, since this sampling optimization is part of the offline training module, it does not affect the acquisition and sampling prediction at test time, which is 0.85 seconds in our experiments.

\section{Conclusion}\label{sec:conclusion}
In this work, we proposed a novel MRI sampling prediction algorithm for multi-coil MRI that estimates a collection of scan-adaptive sampling patterns and a reconstruction network trained on those patterns alternatingly, at training time. {The proposed algorithm was validated on the publicly available fastMRI knee and brain datasets and demonstrated better reconstruction accuracy than the currently used Cartesian undersampling patterns.} 
This study demonstrated the advantages of employing scan-adaptive masks by providing evidence that they are more effectively tailored to individual patients than population-adaptive masks.
We also showed the dependence of the learned sampling patterns on acceleration factors, the initialization of the sampling algorithm, and the reconstruction method used. Future work will include employing deep image prior or other scan-adaptive MRI reconstructions in our framework, extending the approach to cardiac MRI, and/or predicting sparse views for X-ray CT reconstruction.

\section{Acknowledgments}
The authors would like to acknowledge Dr. Maryam Sayadi, Michigan State University
for her inputs throughout the project.
The authors also acknowledge Evan Bell and Shijun Liang from Michigan State University and Zhishen Huang from Amazon Inc. for useful discussions.


\bibliographystyle{IEEEbib}
\bibliography{references}

\end{document}